\documentclass[12pt]{iopart}

\usepackage[]{graphicx}
\eqnobysec

\begin{document} 
\title{Vacuum fluctuations and the conditional homodyne detection of squeezed light} 

\author{H J Carmichael and Hyunchul Nha}
\address{University of Auckland, Private Bag 92019, Auckland, New Zealand}

\ead{h.carmichael@auckland.ac.nz}
 
\begin{abstract}
Conditional homodyne detection of quadrature squeezing is compared with standard
nonconditional detection. Whereas the latter identifies nonclassicality in a
quantitative way, as a reduction of the noise power below the shot noise level,
conditional detection makes a qualitative distinction between vacuum state
squeezing and squeezed classical noise. Implications of this comparison for the
realistic interpretation of vacuum fluctuations (stochastic electrodynamics) are
discussed.
\end{abstract}


\maketitle 

\section{Introduction}
\label{sec:intro}
The field of quadrature squeezing saw its main period of growth in the 1980's
\cite{jmo87,josa87,ap92}. The topic continues to be important to many areas of
research involving nonclassical light and its applications
\cite{Fabre97,Mlynek97,Furusawa98,Bowen03}. In this paper we discuss one of the
earliest and most fundamental issues addressed in this field: the detection of
quadrature squeezed light and the characterization of squeezing as a nonclassical
effect.

The standard squeezing measurement uses balanced homodyne detection \cite{Yuen83}.
In this scheme, nonclassical squeezing is identified with a reduction of the
measured noise variance below the shot noise level. The noise variance depends on
the phase of the local oscillator, in such a way that the reduction evolves into an
enhancement as the phase is continuously changed. In this standard measurement the
shot noise level is the ``measuring stick'', used to differentiate quantum from
classical squeezing in a quantitative way. Considered for the qualitative response
only, the two types of squeezing appear in the same way. In either case, one
observes a phase-sensitive reduction of the noise variance over a limited bandwidth;
as stated, the distinction made is purely quantitative: a reduction below the
shot noise level is a quantum effect, otherwise the squeezing is classical.

There is nothing particularly curious in this. Other instances exist in
quantum optics where nonclassicality is defined by the violation of a quantitative
bound; one might prefer that the bound be a relative measure---a fringe visibility,
for example; on the other hand, there is no reason to doubt that the shot noise level can
be reliably set.

At a more fundamental level, however, it does seem reasonable to expect that
the difference between quantum and classical noise would amount to something more than
the mere {\it size\/} of a noise variance. If, after all, that is the only distinction,
the natural conclusion is that quantum and classical fluctuations are qualitatively
the same. Classical noise models, such as stochastic electrodynamics, should then be
adequate to account for quantum noise.

We show in this paper that in the case of quadrature squeezing the expected
qualitative difference does exist. It is revealed by a measurement based
on {\it conditional\/} homodyne detection \cite{Carmichael00,Foster00}. Our
proposal is not the first to make a qualitative distinction between classical
and quantum squeezing, since such differences also appear in the response of
atoms illuminated by squeezed light \cite{Dalton99}. In contrast to this
earlier work, however, our proposal makes the distinction at the level of an
elementary squeezing measurement.

In section \ref{sec:squeezing} we review the theoretical treatment of quadrature
squeezing of broadband classical noise and its extension to the quantum case via
the Wigner representation. Section \ref{sec:detection} discusses the detection of
squeezed light, and contrasts nonconditional and conditional homodyne detection
schemes. The implications of the demonstrated qualitative difference between
quantum and classical squeezing for the realistic interpretation of vacuum
fluctuations (stochastic electrodynamics) is discussed in section
\ref{sec:stochastic_electrodynamics}. Brief conclusions are then drawn in section
\ref{sec:conclusion}.

\section{Quadrature squeezing as a property of the electromagnetic field}
\label{sec:squeezing}
\begin{figure}[t]
\begin{center}
\begin{tabular}{c}
\hskip-2cm\includegraphics[height=4.75cm]{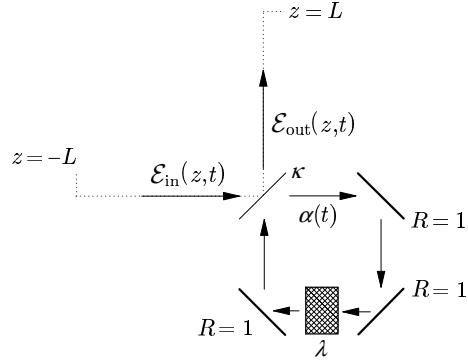}
\end{tabular}
\end{center}
\vskip-0.2in
\caption[example] 
{\label{fig:fig1} 
Sketch of the ring-cavity DPO squeezer with single input-output port. The shaded region
represents the $\chi^{(2)}$ nonlinear crystal.}
\end{figure}

Quadrature squeezing is commonly understood as a property of the electromagnetic field,
without referring to how that property will be measured. This way of thinking is
certainly unproblematic so far as squeezed classical noise is concerned. We therefore
begin from this point of view and review the well-known results for the quadrature
squeezing of a broadband classical noise field, ${\cal E}_{\rm in}(z,t)$ in
figure~\ref{fig:fig1}, injected into the input of a below-threshold degenerate parametric
oscillator (DPO). The squeezing observed at the output is readily understood by considering
how ${\cal E}_{\rm out}(z,t)$ is formed from the interference of ${\cal E}_{\rm in}(z,t)$
with the partial transmission, through the output mirror, of the intracavity field
$\alpha(t)$ \cite{Yurke84}. The spectrum of the squeezing may be calculated in the
following straightforward way.

\subsection{Squeezed classical noise} 
\label{sec:scn}
We first expand the input and output fields in terms of traveling-wave modes satisfying
periodic boundary conditions on an interval of length $2L$ (figure~\ref{fig:fig1}). Mode
frequencies are denoted by $\omega+\omega_0$, where $\omega_0$ is the resonance
frequency of the DPO cavity. The fields, in photon flux units, are then given by
\begin{eqnarray}
{\cal E}_{\rm in}(z,t)&=&\sum_\omega\sqrt{c/2L}\mkern2mu f_\omega
\exp[-\rmi(\omega_0+\omega)(t-z/c)],\label{eqn:input}\\
{\cal E}_{\rm out}(z,t)&=&\sum_\omega\left(\mkern-2mu\sqrt{c/2L}\mkern2mu
f_\omega+\sqrt{2\kappa}\mkern2mu\alpha_\omega\right)\exp[-\rmi(\omega_0+\omega)(t-z/c)],
\label{eqn:output}
\end{eqnarray}
where the complex amplitudes $f_\omega$ are random variables, of zero mean, and
with covariances
\begin{equation}
\overline{f_\omega f_{\omega^\prime}}=\overline{f_\omega^* f_{\omega^\prime}^*}=0,
\qquad\overline{f_\omega^*f_{\omega^\prime}}=\bar n\delta_{\omega\omega^\prime};
\label{eqn:variances}
\end{equation}
the input-field noise spectrum is thus assumed to be flat with an average strength of
$\bar n$ photons per mode. The intracavity field is expanded in a similar way as 
\begin{equation}
\alpha(t)=\sum_\omega\alpha_\omega\rme^{-\rmi(\omega_0+\omega)t},
\end{equation}
and the advertised interference of ${\cal E}_{\rm in}(z,t)$ with $\alpha(t)$ appears
explicitly as a sum of amplitudes $f_\omega$ and $\alpha_\omega$ on the right-hand side
of equation (\ref{eqn:output}). We also introduce the two-mode quadrature amplitudes
\cite{Caves85,Schumaker85} 
\begin{eqnarray} \eqalign{
X_\omega&\equiv(f_\omega+f_{-\omega}^*)/2,\cr
\noalign{\vskip2pt}
Y_\omega&\equiv-\rmi(f_\omega-f_{-\omega}^*)/2,
}\qquad
\eqalign{
x_\omega&\equiv(\alpha_\omega+\alpha_{-\omega}^*)/2,\cr
\noalign{\vskip2pt}
y_\omega&\equiv-\rmi(\alpha_\omega-\alpha_{-\omega}^*)/2.
}
\end{eqnarray}
Note that these are {\it complex\/} quadrature amplitudes and correspond to
{\it non-Hermitian} quantum operators (except when $\omega=0$).

The root cause of squeezing is the phase-sensitive amplification and deamplification of
the intracavity field. It arises because the amplitude $\alpha_\omega$ couples to
$\alpha_{-\omega}^*$ through the $\chi^{(2)}$ nonlinearity of the intracavity crystal.
In addition, $\alpha_\omega$ is excited by $f_\omega$ and damped at the rate $\kappa$.
The pair of coupled amplitude equations are
\begin{eqnarray}
0&=&-(\kappa-\rmi\omega)\alpha_\omega+\kappa\lambda\alpha_{-\omega}^*-\sqrt{c/2L}
\sqrt{2\kappa}\mkern2mu f_\omega,
\label{eqn:plusomega}\\
\noalign{\vskip2pt}
0&=&-(\kappa-\rmi\omega)\alpha_{-\omega}^*+\kappa\lambda\alpha_\omega-\sqrt{c/2L}
\sqrt{2\kappa}\mkern2mu f_{-\omega}^*,
\label{eqn:minusomega}
\end{eqnarray}
where $\lambda$ is a (real) parameter proportional to $\chi^{(2)}$ and the amplitude
of the field that pumps the nonlinear crystal; the threshold for sustained oscillation
of the DPO occurs at $\lambda=1$. The solution of equations (\ref{eqn:plusomega}) and
(\ref{eqn:minusomega}) for the frequency-dependent intracavity quadrature amplitudes is
\begin{eqnarray}
x_\omega&=&-\sqrt{\frac c{2L}}\sqrt{2\kappa}\mkern2mu\frac{X_\omega}{\kappa(1-\lambda)
-\rmi\omega},\label{eqn:xamp}\\
\noalign{\vskip2pt}
y_\omega&=&-\sqrt{\frac c{2L}}\sqrt{2\kappa}\mkern2mu\frac{Y_\omega}{\kappa(1+\lambda)
-\rmi\omega},
\label{eqn:yamp}
\end{eqnarray}
which when substituted into equation (\ref{eqn:output}) yield the relationship between
output- and input-field quadrature amplitudes. The spectra of squeezing are thus
given by the intensities
\begin{equation}
\frac{2L}c\overline{\left|\sqrt{c/2L}\mkern2mu X_\omega+\sqrt{2\kappa}\mkern2mu
x_\omega\right|^2}=\bar n\frac12\frac{[\kappa(1+\lambda)]^2+\omega^2}{[\kappa
(1-\lambda)]^2+\omega^2},
\label{eqn:Xspectrum}
\end{equation}
and
\begin{equation}
\frac{2L}c\overline{\left|\sqrt{c/2L}\mkern2mu Y_\omega+\sqrt{2\kappa}\mkern2mu
y_\omega\right|^2}=\bar n\frac12\frac{[\kappa(1-\lambda)]^2+\omega^2}{[\kappa
(1+\lambda)]^2+\omega^2}.
\label{eqn:Yspectrum}
\end{equation}

The $Y$-quadrature spectrum exhibits squeezing over a bandwidth $2\kappa$. Asymptotically
($\omega\to\pm\infty$) the spectrum is flat, following the spectrum of the input noise.
Squeezing appears as a Lorentzian dip centered at $\omega=0$. The degree of squeezing
increases with the pump parameter $\lambda$, and the Lorentzian dip goes all the way to
zero for $\lambda=1$, corresponding to perfect squeezing on resonance; thus, on resonance
and for $\lambda=1$ the interference of ${\cal E}_{\rm in}(z,t)$ and $\alpha(t)$ produces a
complete cancellation of the input noise---from equation (9), $\sqrt{c/2L}\mkern2mu
Y_{\omega=0}+\sqrt{2\kappa}\mkern2muy_{\omega=0}=0$. The squeezing mechanism
in this classical calculation is remarkably simple and transparent.

\subsection{Vacuum state squeezing in the Wigner representation}
\label{sec:vsswr}
The calculation is readily extended to account for vacuum state squeezing. With
the field quantized, the amplitudes $f_\omega$ and $\alpha(t)$ may be interpreted as
complex amplitudes within the Wigner representation. The quadrature variances then
correspond to operator averages in symmetric order \cite{Carmichael99}, which
requires the substitution
\begin{eqnarray}
4\overline{\left|X_\omega\right|^2}&\to&4\left(\mkern3mu\overline{\left|X_\omega\right|^2}
\mkern3.5mu\right)_W\\\nonumber
&=&\frac12\left\langle(\hat f_\omega^\dagger+\hat f_{-\omega})
(\hat f_\omega+\hat f_{-\omega}^\dagger)+(\hat f_\omega+\hat f_{-\omega}^\dagger)
(\hat f_\omega^\dagger+\hat f_{-\omega})\right\rangle\nonumber\\
&=&2\bar n+1,
\end{eqnarray}
where $\hat f_\omega$ and $\hat f_\omega^\dagger$ are input-mode annihilation and creation
operators. Similarly,
\begin{equation}
4\overline{\left|Y_\omega\right|^2}\to4\left(\mkern3mu\overline{\left|Y_\omega\right|^2}
\mkern3.5mu\right)_W=2\bar n+1.
\end{equation}
The spectra of squeezing become
\begin{equation}
\fl\frac{2L}c\left(\mkern3mu\overline{\left|\sqrt{c/2L}\mkern2mu X_\omega+\sqrt{2\kappa}
\mkern2mu x_\omega\right|^2}\mkern4mu\right)_{\mkern-2mu W}=\left(\bar n+\frac12\right)
\mkern-2mu\frac12\frac{[\kappa(1+\lambda)]^2+\omega^2}{[\kappa(1-\lambda)]^2+\omega^2},
\label{eqn:newXspectrum}
\end{equation}
and
\begin{equation}
\fl\frac{2L}c\left(\mkern3mu\overline{\left|\sqrt{c/2L}\mkern2mu Y_\omega+\sqrt{2\kappa}
\mkern2mu y_\omega\right|^2}\mkern4mu\right)_{\mkern-2mu W}=\left(\bar n+\frac12\right)
\mkern-2mu\frac12\frac{[\kappa(1-\lambda)]^2+\omega^2}{[\kappa(1+\lambda)]^2+\omega^2}.
\label{eqn:newYspectrum}
\end{equation}

Nothing substantial has changed in the calculation of these spectra, nor in the
interpretation of the squeezing mechanism. The only change is an additional noise variance
per mode, $\frac12\langle\hat f_{\omega}^\dagger\hat f_{\omega}+\hat f_{\omega}
\hat f_{\omega}^\dagger\rangle_{\rm vac}=\frac12$. Squeezing is now judged to be quantum
when the Lorentzian dip drops below this vacuum fluctuation level---by the condition
\begin{equation}
\frac{2L}c\left(\mkern3mu\overline{\left|\sqrt{c/2L}\mkern2mu Y_{\omega=0}+\sqrt{2\kappa}
\mkern2mu y_{\omega=0}\right|^2}\mkern4mu\right)_{\mkern-2mu W}<\frac14.
\end{equation}
The on-resonance squeezing is still perfect for $\lambda=1$.

There appears from this calculation to be no physical difference between vacuum
fluctuations and classical noise. Certainly, vacuum fluctuations have their characteristic
strength; but they are not distinguished from classical noise in a qualitative way. For all
practical purposes, vacuum state squeezing is accounted for in the replacement
\begin{equation}
\overline{f_\omega^*f_{\omega^\prime}}=\bar n\delta_{\omega\omega^\prime}\to
\left(\bar n+{\textstyle\frac12\displaystyle}\right)\delta_{\omega\omega^\prime}.
\label{eqn:newvariances}
\end{equation}
One is tempted to view the vacuum field as a visualizable reality, no less real than the
classical noise itself. Then quadrature squeezing appears to make the case of stochastic
electrodynamics \cite{Marshall63,Boyer75,Marshall88,Casado97a,Casado97b,Casado97c,Casado98}:
the vacuum field is real, merely a {\it stochastic\/} component of the (asymptotic)
input Maxwell field.

We know, however, that there {\it is\/} a physical distinction between vacuum fluctuations
and classical noise: a classical noise field causes a photodetector to fire; vacuum
fluctuations, on the other hand, do not. The difference is important where the measurement
of squeezing is concerned, and particularly so when conditional homodyne detection is used,
since in this case the data taking is triggered by a photocount
\cite{Carmichael00,Foster00}. Indeed, as we aim to show, through conditional homodyne
detection vacuum state squeezing is revealed to be a qualitatively distinct phenomenon
from the squeezing of classical noise. In order to demonstrate how and why, we turn our
attention now to the detection of squeezed light.

\section{Quadrature squeezing as a scattering process between classical inputs and outputs}
\label{sec:detection}
Experiments in quantum optics begin with inputs that can be described in classical terms
and end with classical data records---time series of real numbers. In this sense, they are
scattering processes between classical inputs and outputs. In some instances a classical
model serves to map the inputs to outputs. More generally, quantum mechanical ideas
are needed for a correct and consistent account. Squeezing, viewed as a scattering process,
is a process of the latter sort. The classical inputs in the model of figure~\ref{fig:fig1}
are the noise field ${\cal E}_{\rm in}(z,t)$ and the DPO pump field, represented by the
parameter~$\lambda$. Stochastic electrodynamics would have us add a realistic vacuum field
to this, and thus view squeezing as a scattering process of the classical type. Conventional
opinion regards such a view to be problematic, in spite of the calculation
leading from equation (\ref{eqn:newvariances}) to equations (\ref{eqn:newXspectrum}) and
(\ref{eqn:newYspectrum}). Problems arise with the generation of data records through the
process of photoelectric detection; inevitable difficulties and inconsistencies appear
such as vacuum-induced firings of the detectors. These problems are noted as a secondary
theme in what follows. Our primary interest, however, is with the comparison between standard
nonconditional and conditional homodyne detection. In particular, the demonstration of how
the latter distinguishes quantum squeezing from its classical counterpart.

\subsection{Homodyne detection for classical fields}
\label{sec:hdcf} 
\begin{figure}[t]
\begin{center}
\begin{tabular}{c}
\raise0.15in\hbox{\includegraphics[height=3.75cm]{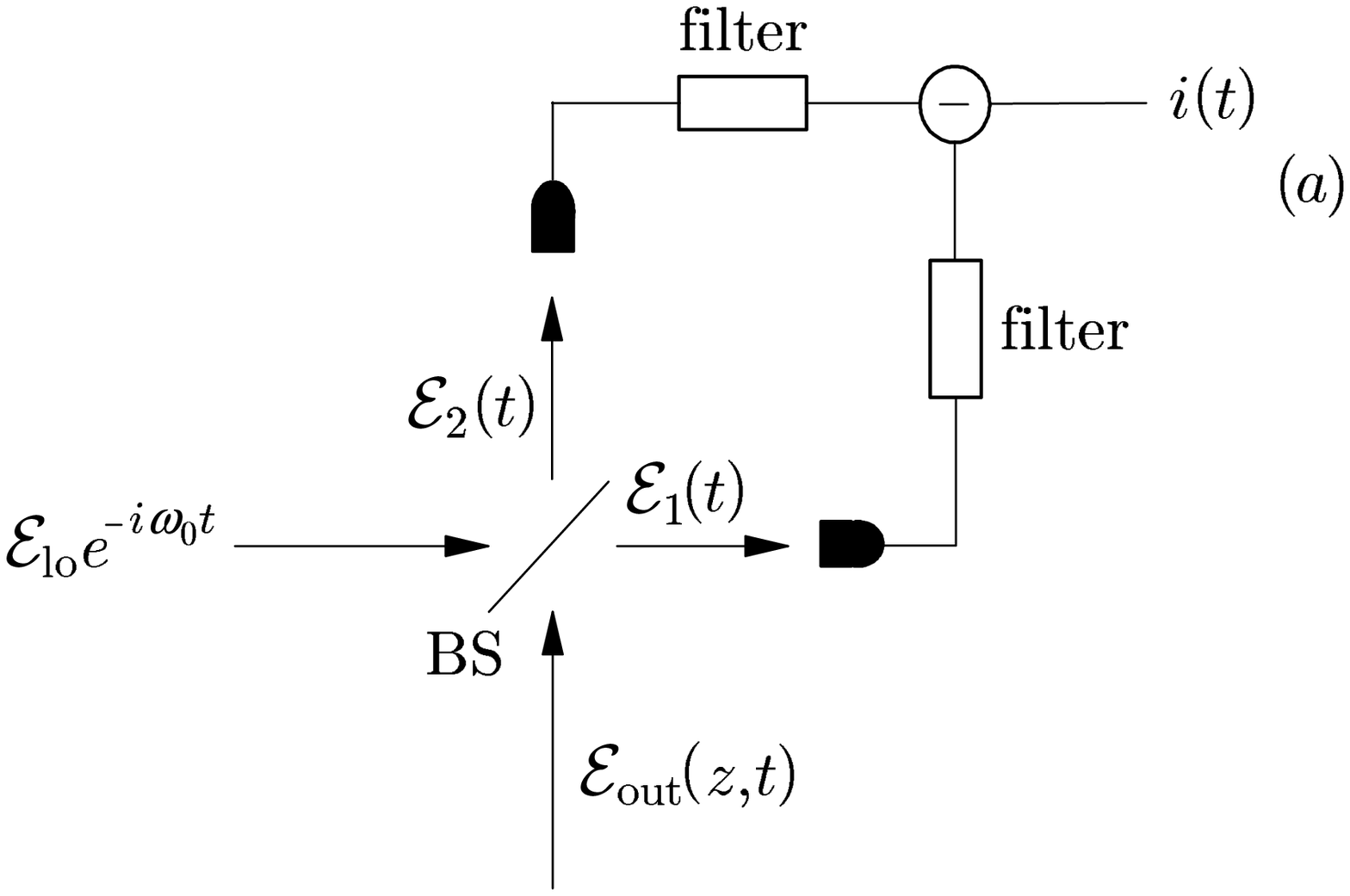}}
\hskip.75cm
\includegraphics[height=3.9cm]{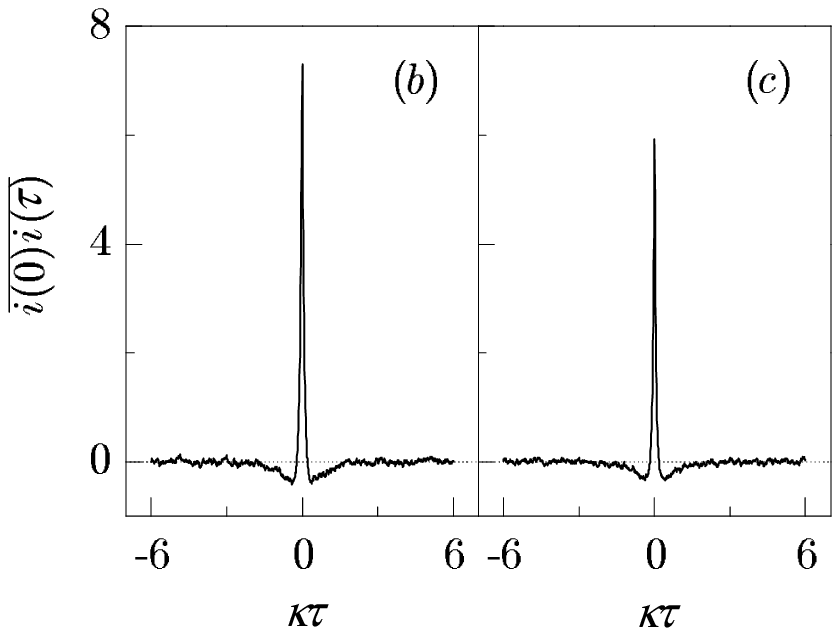}
\end{tabular}
\end{center}
\vskip-0.2in
\caption[example] 
{\label{fig:fig2} 
({\it a}) In balanced homodyne detection the signal field ${\cal E}_{\rm out}(t)$ is
mixed with a local oscillator at a 50/50 beam splitter (BS). Photoelectric detection of
the outputs generates two photocurrents, which are filtered and subtracted
to produce the current $i(t)$. ({\it b}) and ({\it c}) Autocorrelation of $i(t)$ reveals
nonsqueezed fluctuations over a wide bandwidth (central spike)
and a narrow bandwidth of squeezed fluctuations (negative dip). The plotted correlation
functions are calculated for the model of figure~\ref{fig:fig4}, with classical noise
bandwidth $B_{\rm c}=15$, detection bandwidth $B_{\rm d}=25$, DPO pump parameter
$\lambda=0.4$, and classical noise photon number in the DPO cavity $\bar n_a=0.2$
({\it b}) and $\bar n_a=0$ ({\it c}).}
\end{figure}

Unconditional balanced homodyne detection of the classical field ${\cal E}_{\rm out}(z,t)$
is accomplished using the scheme sketched in figure~\ref{fig:fig2}({\it a}) \cite{Yuen83}.
The output data record is the current $i(t)$. Its generation from ${\cal E}_{\rm out}(z,t)$
follows in an elementary way. On combining ${\cal E}_{\rm out}(z,t)$ with a strong local
oscillator field, amplitude ${\cal E}_{\rm lo}$, at a 50/50 beam splitter, the two fields
\begin{eqnarray}
{\cal E}_1(t)&=&[{\cal E}_{\rm lo}\exp(-\rmi\omega_0t)+{\cal E}_{\rm out}(t)]/\sqrt2
\mkern2mu,\\
\noalign{\vskip2pt}
{\cal E}_2(t)&=&[{\cal E}_{\rm lo}\exp(-\rmi\omega_0t)-{\cal E}_{\rm out}(t)]/\sqrt2
\mkern2mu,
\end{eqnarray}
are produced, with ${\cal E}_{\rm out}(t)\equiv{\cal E}_{\rm out}(z_o,t)$, $z_o$ some
arbitrary location. Their intensities are
\begin{eqnarray}
|{\cal E}_1(t)|^2&\approx&{\textstyle\frac12\displaystyle}|{\cal E}_{\rm lo}|^2
+{\textstyle\frac12\displaystyle}|{\cal E_{\rm lo}}|
[\exp(-\rmi\phi_{\rm lo})\tilde{\cal E}_{\rm out}(t)+\exp(\rmi\phi_{\rm lo})
\tilde{\cal E}_{\rm out}^*(t)],
\label{eqn:intensity1}\\
\noalign{\vskip3pt}
|{\cal E}_2(t)|^2&\approx&{\textstyle\frac12\displaystyle}|{\cal E}_{\rm lo}|^2
-{\textstyle\frac12\displaystyle}|{\cal E_{\rm lo}}|
[\exp(-\rmi\phi_{\rm lo})\tilde{\cal E}_{\rm out}(t)+\exp(\rmi\phi_{\rm lo})
\tilde{\cal E}_{\rm out}^*(t)],
\label{eqn:intensity2}
\end{eqnarray}
where small terms $|\tilde{\cal E}_1(t)|^2$ and $|\tilde{\cal E}_2(t)|^2$ are neglected,
$\phi_{\rm lo}$ is the local oscillator phase, and
\begin{equation}
\tilde{\cal E}_{\rm out}(t)=\exp(\rmi\omega_0 t){\cal E}_{\rm out}(t).
\label{eqn:tildeout}
\end{equation}
The intensities $|{\cal E}_1(t)|^2$ and $|{\cal E}_2(t)|^2$ determine the rates of
photoelectron generation at the detectors. Since photoelectron emission is a random process,
the difference ``current'' (units $[{\rm time}]^{-1/2}$) satisfies a stochastic differential
equation,
\begin{equation}
\rmd i=-B_{\rm d}\kappa(i\rmd t-\rmd Q),
\label{eqn:homocurrent}
\end{equation}
where  $2B_{\rm d}\kappa$ is the detection bandwidth and
\begin{equation}
\rmd Q=2\tilde{\cal E}_{\rm out}^{\rm Y}(t)\rmd t+\rmd W_t
\label{eqn:homocharge}
\end{equation}
is the ``charge'' (units $[{\rm time}]^{1/2}$) deposited in time step $dt$, with $dW_t$ a
Weiner increment ($\overline{dW_tdW_t}=dt$) introduced to account for the Poisson fluctuation
in photoelectron number (shot noise). For an appropriate choice of $\phi_{\rm lo}$,  
\begin{equation}
\tilde{\cal E}_{\rm out}^{\rm Y}(t)=\sqrt{c/2L}\mkern2mu Y(t)+\sqrt{2\kappa}
\mkern2mu y(t),
\label{eqn:homoamplitude}
\end{equation}
with $Y(t)=\sum_\omega Y_\omega\rme^{-\rmi\omega t}$ and similarly for $y(t)$, and the
current $i(t)$ records a filtered version of the squeezed quadrature amplitude $\sqrt{c/2L}
\mkern2mu Y(t)+\sqrt{2\kappa}\mkern2mu y(t)$, contaminated by shot noise $W_t=\int_0^t\rmd
W_{t^\prime}$. 

We might propose an extension of this treatment to vacuum state squeezing in the manner
of section~\ref{sec:vsswr}, simply increasing the input noise variances as in equation
(\ref{eqn:newvariances}). The strategy  meets with a difficulty here, though, since
$\rmd Q$ already includes shot noise; adding realistic vacuum fluctuations double-counts
this noise. From an operational point of view, one might certainly remove the
$\rmd W_t$ from  equation (\ref{eqn:homocharge}), add vacuum fluctuations to $Y(t)$,
and effectively move the shot noise into the signal $\tilde{\cal E}_{\rm out}^{\rm Y}(t)$.
The result is a rather unsatisfactory modeling of photoelectric detection, where the
photocurrent is identified {\it deterministically\/} with the fluctuating light intensity;
the production of photoelectrons is not longer a random process. Rather than dwell on
this issue, we switch to a quantum mechanical treatment of homodyne detection.

\subsection{Homodyne detection in quantum trajectory theory}
\label{sec:hdqtt}
Quantum trajectory theory \cite{Carmichael93b} makes only a small change to the scheme set
out in equations (\ref{eqn:homocurrent}), (\ref{eqn:homocharge}), and \ref{eqn:homoamplitude}).
In equation (\ref{eqn:homoamplitude}) the replacement
\begin{equation}
y(t)\to\langle\tilde\psi_{W_t}(t)|\hat y|\tilde\psi_{W_t}(t)\rangle
\end{equation}
is made, where $\hat y$ is the operator quadrature
amplitude
\begin{equation}
\hat y\equiv-\rmi(\hat{\tilde a}-\hat{\tilde a}^\dagger)/2,\qquad \hat{\tilde a}
\equiv\exp(\rmi\omega_0t)\hat a,
\end{equation}
and $|\tilde\psi_{W_t}(t)\rangle\equiv|\tilde{\bar\psi\mkern4mu}\mkern-4mu_{W_t}(t)\rangle
/\big[\langle\tilde{\bar\psi\mkern4mu}\mkern-4mu_{W_t}(t)|\tilde{\bar\psi\mkern4mu}
\mkern-4mu_{W_t}(t)\rangle\big]^{1/2}$ is the quantum state conditioned on a realization of
the shot noise $W_t$; for the model of figure~\ref{fig:fig1} it satisfies the stochastic
Schr\"odinger equation
\begin{eqnarray}
\fl
\rmd|\tilde{\bar\psi\mkern4mu}\mkern-4mu_{W_t}\rangle=\left\{\mkern-2mu\left[-\kappa
\hat{\tilde a}^\dagger\hat{\tilde a}+(\kappa\lambda/2)(\hat{\tilde a}^{\dagger2}-
\hat{\tilde a}^2)\right.\right.\nonumber\\
\left.\left.\vphantom{\hat{\tilde a}^\dagger}-\sqrt{2\kappa}
\left[\vphantom{\hat a^\dagger}\right.\mkern-2mu\tilde{\cal E}_{\rm in}(t)\hat a^\dagger
-\tilde{\cal E}_{\rm in}^*(t)\hat a\mkern-2mu\left.\vphantom{\hat a^\dagger}\right]
\right]\mkern-3mu\rmd t-\rmi\sqrt{2\kappa}\mkern2mu\hat{\tilde a}\rmd Q\right\}
\mkern-2mu|\tilde{\bar\psi\mkern4mu}\mkern-4mu_{W_t}\rangle,
\label{eqn:stochschroedinger}
\end{eqnarray}
with
\begin{equation}
\tilde{\cal E}_{\rm in}(t)=\exp(\rmi\omega_0t){\cal E}_{\rm in}(z_i,t),
\end{equation}
$z_i$ an arbitrary location.

Figures~\ref{fig:fig2}({\it b}) and ({\it c}) illustrate how the squeezing of equation
(\ref{eqn:newYspectrum}) appears in standard homodyne detection. The simulations are
based upon the model depicted in figure~\ref{fig:fig4} of a finite bandwidth input, where
equation (\ref{eqn:newYspectrum}) has been generalized to a frequency-dependent
variance [$\bar n\to\bar n(\omega)$]. Results are presented in the time domain; we
plot the autocorrelation function of the homodyne current $\overline{i(0)i(\tau)}$---i.e.,
the Fourier transform of the output current power spectrum. The correlation function is
the sum of two terms: a tall narrow spike, arising from the broad background of
nonsqueezed fluctuations in the frequency domain, and a wider, negative dip, corresponding
to the finite bandwidth of quadrature squeezing. We are interested specifically in the
narrow spike, which in turn is the sum of two contributions: the nonsqueezed shot noise
(vacuum fluctuations)---the $1/2$ of $[\bar n(\omega)+1/2]$ in equation
(\ref{eqn:newYspectrum})---and the nonsqueezed classical noise described by the $\bar n
(\omega)$, both filtered through the detection bandwidth. The classical noise strength
is zero in figure~\ref{fig:fig2}({\it c}) and hence the height of the spike is reduced
in comparison with figure~\ref{fig:fig2}({\it b}).

Our main result, demonstrated in the section~\ref{sec:vssvscn}, is that conditional
homodyne detection differentiates between the two nonsqueezed contributions. It eliminates
the shot noise and retains only the classical noise; the central spike disappears
altogether in the case of vacuum state squeezing [figure~\ref{fig:fig5}(d) versus
 \ref{fig:fig2}({\it c})].

\subsection{Conditional homodyne detection}
\label{sec:chd}
\begin{figure}[b]
\begin{center}
\begin{tabular}{c}
\includegraphics[height=6.25cm]{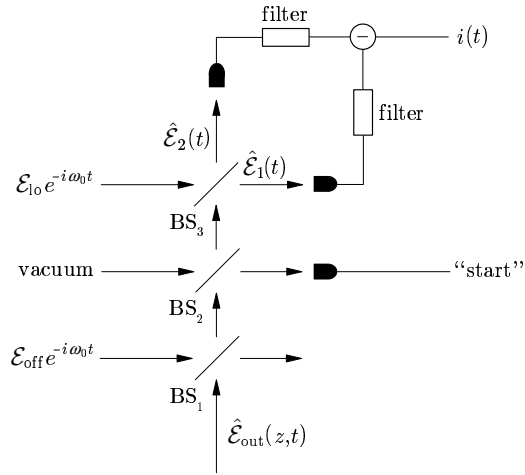}
\end{tabular}
\end{center}
\vskip-0.2in
\caption[example] 
{\label{fig:fig3} 
In conditional homodyne detection, the signal field $\hat{\cal E}_{\rm out}(t)$ is
displaced by a weak coherent offset ${\cal E}_{\rm off}$ and sampling of the 
current $i(t)$ is triggered by a ``start'' photocount.
Beam splitters ${\rm BS}_1$ and ${\rm BS}_2$ can have arbitrary reflection and
transmission. ${\rm BS}_3$ is a 50/50 beam splitter.}
\end{figure}

The conditional homodyne measurement scheme is sketched in figure~\ref{fig:fig3}. The idea
is to sample the current $i(t+\tau)$ only when the ``start'' photodetector has fired (will
fire) at time~$t$. The average over many samples yields a conditional average of $i(t+\tau)$.
The scheme measures the cross-correlation of the light intensity in the ``start'' channel
and the quadrature amplitude selected by the homodyne detector. For a squeezing measurement,
correct settings of the phases of the coherent offset, ${\cal E}_{\rm off}
\exp(-\rmi\omega_0t)$, and the local oscillator are required, and the choice of the amplitude
$|{\cal E}_{\rm off}|$ affects the normalization of the measured correlation function.
These details are discussed elsewhere \cite{Carmichael00,Foster00,Foster02}. Assuming an
optimum choice for  ${\cal E}_{\rm off}$, a detection bandwidth much larger than the signal
bandwidth, and Gaussian fuctuations, the measured correlation function is
\begin{equation}
h_{\rm Y}(\tau)=1+\frac{\langle\mkern2mu:\hat{\tilde{\cal E}\mkern4mu}\mkern-4mu
{}_{\rm out}^{\rm Y}(0)\hat{\tilde{\cal E}\mkern4mu}\mkern-4mu{}_{\rm out}^{\rm Y}
(\tau)\mkern-3mu:\mkern2mu\rangle}{\langle\hat{\tilde{\cal E}\mkern4mu}
\mkern-4mu{}_{\rm out}^{\dagger}(0)\hat{\tilde{\cal E}\mkern4mu}\mkern-4mu{}_{\rm out}
(0)\rangle},
\label{eqn:htau}
\end{equation}
where $\langle\mkern2mu:\mkern3mu:\mkern2mu\rangle$ denotes the normal- and time-ordered
quantum average.

In contrast to the autocorrelation of $i(t)$ plotted in figure~\ref{fig:fig2}, the
time-displaced quadrature amplitudes entering this expression, $\hat{\tilde{\cal E}\mkern4mu}
\mkern-4mu{}_{\rm out}^{\rm Y}(0)$ and $\hat{\tilde{\cal E}\mkern4mu}\mkern-4mu{}_{\rm out}
^{\rm Y}(\tau)$, have physically different origins: the first comes from the interference of
$\hat{\cal E}_{\rm out}(t)$ with ${\cal E}_{\rm off}\exp(-\rmi\omega_0t)$ in the square law
response of the ``start'' photodetector; the second from the interference of $\hat
{\cal E}_{\rm out}(t+\tau)$ and ${\cal E}_{\rm lo}\exp[-\rmi\omega_0t+\tau)]$ at beam
splitter ${\rm BS}_3$. Thus, in place of the autocorrelation plotted in figure~\ref{fig:fig2},
we have a cross-correlation of the ``start'' channel with $i(t)$. From this it follows that
if the shot noise enters the homodyne current due to the randomness of photoelectric detection,
rather than as a real fluctuation carried by $\hat{\cal E}_{\rm out}(t)$, then $h(\tau)$
should not include a contribution from the autocorrelation of the nonsqueezed shot noise,
though the classical-noise autocorrelation should still be present. The distinction is made,
mathematically, by the normal ordering of the quantum average.

\subsection{Vacuum state squeezing versus squeezed classical noise} 
\label{sec:vssvscn}
\begin{figure}[b]
\begin{center}
\begin{tabular}{c}
\hskip-0.75cm\includegraphics[height=4.25cm]{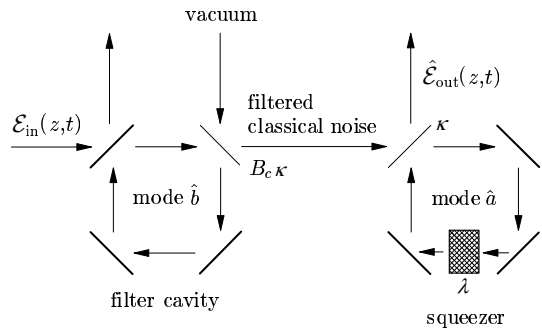}
\end{tabular}
\end{center}
\vskip-0.2in
\caption[example] 
{\label{fig:fig4} 
Sketch of the finite bandwidth classical noise source and squeezer.}
\end{figure}

To demonstrate this prediction we have simulated conditional homodyne detection for
the model of figure~\ref{fig:fig4}. In the model, the broadband classical noise field
${\cal E}_{\rm in}(z,t)$ is passed through a filter cavity to produce noise of
bandwidth $2B_{\rm c}\kappa$. The filtered noise provides the input to the DPO
squeezer, whose squeezed output, $\hat{\cal E}_{\rm out}(z,t)$, is fed to the
conditional homodyne detector. The entire system may be viewed as a scattering process
from the classical inputs ${\cal E}_{\rm in}(z,t)$, $\lambda$, ${\cal E}_{\rm off}
\exp(-\rmi\omega_0t)$, and ${\cal E}_{\rm lo}\exp(-\rmi\omega_0t)$, to classical data
records composed of the ``start'' counts and $i(t)$. The simulations were in fact
performed using the quantum trajectory theory of cascaded open systems \cite{Carmichael93a},
by treating both cavity modes as quantized fields as indicated by the operator labels in
figure~\ref{fig:fig4}. Equally well, the squeezer might be provided with a classical
input field $\tilde{\cal E}_{\rm in}^\prime(t)$ (in photon flux units) that satisfies the
stochastic differential equation
\begin{equation}
\rmd\tilde{\cal E}_{\rm in}^\prime=-B_{\rm c}\kappa\tilde {\cal E}_{\rm in}^\prime\rmd
t-2\sqrt{B_{\rm c}\kappa\kappa^\prime\bar n}\mkern2mu\rmd W_t^\prime,\qquad B_{\rm c}\kappa\gg
\kappa^\prime, 
\end{equation}
where $\rmd W^\prime_t/\rmd t$ models the filter cavity input, with
$\rmd W_t^\prime$ a Weiner increment.

\begin{figure}[t]
\vskip0.1in
\begin{center}
\begin{tabular}{c}
\includegraphics[height=3.75cm]{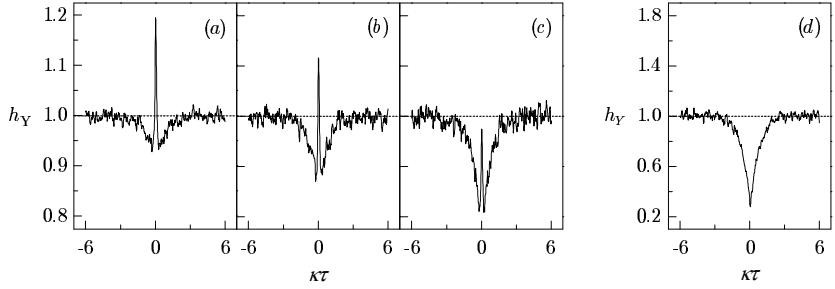} 
\end{tabular}
\end{center}
\vskip-0.2in
\caption[example] 
{\label{fig:fig5} 
The correlation function $h_{\rm Y}(\tau)$ calculated for the model of figure~\ref{fig:fig4}:
with classical noise bandwidth $B_{\rm c}=15$, detection bandwidth $B_{\rm d}=25$, DPO
pump parameter $\lambda=0.4$, and classical noise photon number in the DPO cavity $\bar n_a
=0.2$ ({\it a}), $\bar n_a=0.1$ ({\it b}), $\bar n_a=0.05$ ({\it c}), and $\bar n_a=0$
({\it d}).}
\end{figure}

Figure~\ref{fig:fig5} presents results from the simulated measurement of
$h_{\rm Y}(\tau)$, where figures \ref{fig:fig5}({\it a}) and ({\it d}) correspond to
figures~\ref{fig:fig2}({\it b}) and ({\it c}), respectively. The shot noise contribution
to the central spike has disappeared while the contribution from the classical
noise remains; for vacuum state squeezing there is no spike at all. Thus, vacuum state
squeezing is distinguished {\it qualitatively\/} by the measurement from squeezed classical
noise.

\section{Conditional homodyne detection in stochastic electrodynamics}
\label{sec:stochastic_electrodynamics}
We turn now to our subsidiary theme.
Given the comparison between figure~\ref{fig:fig5} and figures~\ref{fig:fig2}(b) and (c),
what does conditional homodyne detection have to say about realistic vacuum
fluctuations---about stochastic electrodynamics?

Consider again our argument in section~\ref{sec:chd} for the disappearance of the shot
noise spike. It regards the shot noise to be produced in the photoelectric detection process:
shot noise is not derived from a fluctuation of the field and is therefore distinguishable
from the classical noise which is. It appears then that stochastic electrodynamics
suffers a fatal blow from the demonstrated results, as it sees shot noise precisely as 
additional (indistinguishable) fluctuations added to the field. Such a conclusion is too hasty,
though. Stochastic electrodynamics does a better job of describing conditional homodyne
detection than one might initially expect. Certainly it meets with difficulties of the usual
sort. Nevertheless, it also predicts the vanishing of the shot noise spike; it offers an
entirely different explanation of the effect. We conclude by modeling the conditional
measurement of squeezing within stochastic electrodynamics.

\begin{figure}[t]
\begin{center}
\begin{tabular}{c}
\includegraphics[height=6.5cm]{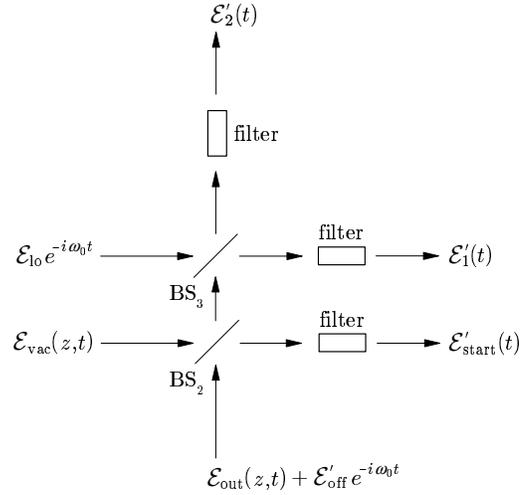}
\end{tabular}
\end{center}
\vskip-0.2in
\caption[example] 
{\label{fig:fig6} 
The field arrangement used to model conditional homodyne detection in stochastic
electrodynamics.}
\end{figure}

The Achilles heel of stochastic electrodynamics is its inability to give a plausible
account of the firing of photoelectric detectors. To avoid this important yet
distracting issue, we set aside a consideration of the detection process itself and
return to the attitude taken in section~\ref{sec:squeezing}; we simply calculate
moments of the measured fields. The arrangement used is illustrated in figure~\ref{fig:fig6}.
There are three fields to consider: 
\begin{eqnarray}
{\cal E}_1(t)&=&\left[{\cal E}_{\rm lo}\exp(-\rmi\omega_0t)+\sqrt{1-r}\mkern2mu
{\cal E}_{\rm out}(t)-\sqrt r\mkern2mu{\cal E}_{\rm vac}(t)\right]/\sqrt2,\\
\noalign{\vskip2pt}
{\cal E}_2(t)&=&\left[{\cal E}_{\rm lo}\exp(-\rmi\omega_0t)-\sqrt{1-r}\mkern2mu
{\cal E}_{\rm out}(t)+\sqrt r\mkern2mu{\cal E}_{\rm vac}(t)\right]/\sqrt2,
\end{eqnarray}
and
\begin{equation}
{\cal E}_{\rm start}(t)=\sqrt{r}\mkern2mu{\cal E}_{\rm out}(t)+\sqrt{1-r}\mkern2mu
{\cal E}_{\rm vac}(t),
\end{equation}
where $r$ is the reflection coefficient of beam splitter ${\rm BS}_2$ (figure~\ref{fig:fig6})
and
\begin{equation}
{\cal E}_{\rm vac}(t)\equiv{\cal E}_{\rm vac}(z_v,t),
\end{equation}
$z_v$ an arbitrary location. Each field is filtered, with bandwidth $B_{\rm d}\kappa$,
so that the vacuum fluctuations have finite photon flux. We calculate the correlation
function as
\begin{equation}
h^\prime_{\rm Y}(\tau)=\frac{\overline{|{\cal E}_{\rm start}^\prime(0)|^2
[|{\cal E}_1^\prime(\tau)|^2-|{\cal E}_2^\prime(\tau)|^2]}}
{\overline{|{\cal E}_{\rm start}^\prime(0)|^2}\times
\overline{[|{\cal E}_1^\prime(0)|^2-|{\cal E}_2^\prime(0)|^2]}};
\label{eqn:hprimetau1}
\end{equation}
${\cal E}_1^\prime(t)$, ${\cal E}_2^\prime(t)$, and ${\cal E}_{\rm start}^\prime(t)$ are
the filtered fields. The question for us now is the following: what does this expression have to say about
the shot noise spike?

In stochastic electrodynamics the signal field ${\cal E}_{\rm out}(t)$
carries realistic vacuum fluctuations, injected at the vacuum input to the filter cavity
(figure~\ref{fig:fig4}). If this were the only vacuum input, the spike would remain in
$h^\prime_{\rm Y}(\tau)$; but conditional detection introduces a second vacuum input---the
field ${\cal E}_{\rm vac}(z,t)$ injected at beam splitter ${\rm BS}_2$. Considering it,
the expansion of equation (\ref{eqn:hprimetau1}) (for Gaussian fluctuations) yields
\begin{equation}
h^\prime_{\rm Y}(\tau)=1+A\left[\overline{\tilde{\cal E}_{\rm out}^{\rm Y}(0)
\tilde{\cal E}_{\rm out}^{\rm Y}(\tau)}-\overline{\tilde{\cal E}_{\rm vac}^{\rm Y}(0)
\tilde{\cal E}_{\rm vac}^{\rm Y}(\tau)}\right],
\label{eqn:hprimetau2}
\end{equation}
where $A$ is a constant that depends on the amplitude of the offset
$|{\cal E}_{\rm off}^\prime|$, the reflection coefficient $r$, and the filter bandwidth
$2B_{\rm d}\kappa$. We see that the shot noise contribution to the spike {\it is\/}
eliminated, not through the normal ordering of quantum operators, but by the explicit
subtraction of the autocorrelation of the realistic vacuum fluctuations fed through
beam splitter ${\rm BS}_2$, the term $-\overline{\tilde{\cal E}_{\rm vac}^{\rm Y}(0)
\tilde{\cal E}_{\rm vac}^{\rm Y}(\tau)}$ \cite{Note}.

\begin{figure}[b]
\begin{center}
\begin{tabular}{c}
\includegraphics[height=3.75cm]{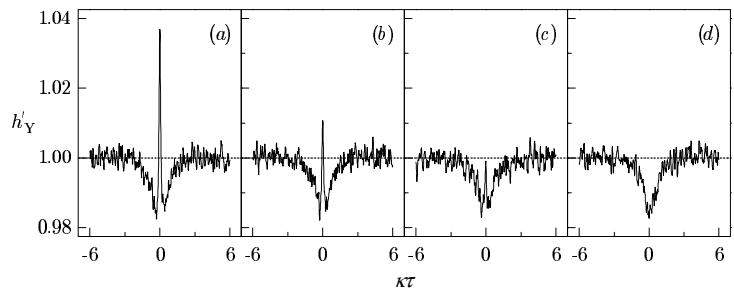}
\end{tabular}
\end{center}
\vskip-0.2in
\caption[example] 
{\label{fig:fig7} 
The correlation function $h^\prime_Y(\tau)$ calculated within stochastic
electro-dynamics: for the parameters of figure~\ref{fig:fig5}.}
\end{figure}
Simulated results for $y_{\rm Y}^\prime(\tau)$ are presented in figure~\ref{fig:fig7}.
They are in qualitative agreement with figure~\ref{fig:fig5}; although all is not well at
a quantitative level; there is a difference in both the absolute and relative sizes of the
squeezing dips. This comes from the inevitable difficulties faced by stochastic
electrodynamics. A problem arises due to the presence in $|{\cal E}_{\rm start}(t)|^2$ of
a nonphysical vacuum field photon flux proportional to the filter bandwidth $2B_{\rm d}
\kappa$---i.e., nonphysical dark counts at the ``start'' detector. If we are prepared
to set this inevitable problem aside, however, stochastic electrodynamics provides an
alternative explanation for the elimination of the shot noise spike.

\section{Conclusions}
\label{sec:conclusion}
We have compared conditional and nonconditional measurements of quadrature
squeezing and shown that conditional detection distinguishes {\it qualitatively\/} between
quantum and classical squeezing. We showed that both measurements may be understood
at a superficial level by adding realistic vacuum fluctuations to asymptotic input fields
(stochastic electrodynamics). The strategy is unable to offer a plausible treatment of
photoelectric detection, however, and this leads to serious quantitative errors
in the treatment of the conditional measurement. 

\ack      
This work was supported by the National Science Foundation under Grant No.\
PHY-0099576.

\end{document}